\documentclass[a4paper,aps,prd,onecolumn,nofootinbib,preprintnumbers,superscriptaddress,amsmath,amssymb]{revtex4-2}

\usepackage{graphicx}
\usepackage{bm,natbib,url,textcase}
\usepackage[english]{babel}
\usepackage[T1]{fontenc}
\usepackage{braket}
\usepackage{quoting}
\quotingsetup{font=small}
\usepackage{csquotes}
\usepackage{dcolumn}
\usepackage{microtype}
\numberwithin{equation}{section}
\usepackage[colorlinks]{hyperref}
\usepackage[dvipsnames,svgnames,x11names]{xcolor}

\usepackage{mathrsfs}

\usepackage{multirow}
\usepackage{epstopdf}
\usepackage{times}

\hypersetup{
	colorlinks=true,                % false: boxed links; true: colored links
	linkcolor=red!60!black,         % color of internal links
	citecolor=green!55!black,       % color of links to bibliography
	urlcolor=blue!40!black,         % color of urls
}

\makeatletter
\renewcommand\@makefnmark{\hbox{\@textsuperscript{\normalfont\color{blue!40!black}\@thefnmark}}}
\makeatother

\usepackage{latexsym}
\def\be{\begin{equation}}
	\def\ee{\end{equation}}
\def\bea{\begin{eqnarray}}
	\def\eea{\end{eqnarray}}
\def\ben{\begin{enumerate}}
	\def\een{\end{enumerate}}

\usepackage{mathtools}
\DeclarePairedDelimiter\abs{\lvert}{\rvert}%
\makeatletter
\let\oldabs\abs
\def\abs{\@ifstar{\oldabs}{\oldabs*}}

\def\a{\alpha}
\def\b{\beta}

\def\c{\rho_{c}}

\def\r{\rho}

\def\k{\kappa}

\def\g{\gamma}
\def\L{\Lambda}\def\l{\lambda}

\def\d{\delta}
\def\p{\partial}

\def\av{``}
\def\cv{''}
\def\z{~}
\def\eps{\epsilon}

\def\nn{\nonumber}
\def\f{\varphi}

\usepackage{tikz}
\definecolor{purple}{rgb}{1,0,1}
\definecolor{lime}{HTML}{A6CE39} % needs xcolor

\newcommand{\orcidicon}{%
	\begin{tikzpicture}
		\draw[lime, fill=lime] (0,0) 
		circle [radius=0.16] 
		node[white] {{\fontfamily{qag}\selectfont \tiny ID}};
		\draw[white, fill=white] (-0.0625,0.095) 
		circle [radius=0.007];
	\end{tikzpicture}	\hspace{-2mm}
}
\newcommand\orcidMarcello{{\href{https://orcid.org/0000-0003-0397-2705}{\orcidicon}}}
\newcommand\orcidDaniele{{\href{https://orcid.org/0000-0003-4379-2549}{\orcidicon}}}
\newcommand\orcidSalvatore{{\href{https://orcid.org/0000-0003-4886-2024}{\orcidicon}}}
\newcommand\orcidFrancisco{{\href{https://orcid.org/0000-0002-9388-8373}{\orcidicon}}}
\begin{document}
    
    \title{Effective Actions for Loop Quantum Cosmology in Fourth-Order Gravity}
	%=================================================================
	
	\author{Marcello Miranda\orcidMarcello\!\!}
	\email{marcello.miranda@unina.it}
	\affiliation{Scuola Superiore Meridionale, Largo San Marcellino 10, I-80138, Napoli, Italy.}
	\affiliation{INFN Sez. di Napoli, Compl. Univ. di Monte S. Angelo, Edificio G, Via	Cinthia, I-80126, Napoli, Italy.}
	
	%=================================================================
	
	\author{Daniele Vernieri\orcidDaniele\!\!}
	\email{daniele.vernieri@unina.it}
	\affiliation{Dipartimento di Fisica ``E. Pancini'', Universit\`{a} di Napoli ``Federico II'', Napoli, Italy.}
	\affiliation{INFN Sez. di Napoli, Compl. Univ. di Monte S. Angelo, Edificio G, Via	Cinthia, I-80126, Napoli, Italy.}
	
	%=================================================================
	
	\author{Salvatore Capozziello\orcidSalvatore\!\!}
	\email{capozziello@na.infn.it}
	\affiliation{Scuola Superiore Meridionale, Largo San Marcellino 10, I-80138, Napoli, Italy.}
	\affiliation{Dipartimento di Fisica ``E. Pancini'', Universit\`{a} di Napoli ``Federico II'', Napoli, Italy.}
	\affiliation{INFN Sez. di Napoli, Compl. Univ. di Monte S. Angelo, Edificio G, Via	Cinthia, I-80126, Napoli, Italy.}
	\affiliation{Laboratory of Theoretical Cosmology, Tomsk State University of Control Systems and Radioelectronics (TUSUR), 634050 Tomsk, Russia.}
	
	%=================================================================
	
	\author{Francisco S. N. Lobo\orcidFrancisco\!\!}
	\email{fslobo@fc.ul.pt}
	\affiliation{Instituto de Astrof\'{i}sica e Ci\^{e}ncias do Espaco, Faculdade de Ci\^encias da Universidade de Lisboa, Edif\'{i}cio C8, Campo Grande, PT-1749-016, Lisbon, Portugal.}
	\affiliation{Departamento de F\'{i}sica, Faculdade de Ci\^{e}ncias, Universidade de Lisboa, Edif\'{i}cio C8, Campo Grande, PT-1749-016 Lisbon, Portugal.}
	
	%=================================================================
	\date{\LaTeX-ed \today}
	%========================================================
	\begin{abstract}
	Loop Quantum Cosmology (LQC) is a theory which renders the Big Bang initial singularity into a quantum bounce, by means of short-range repulsive quantum effects at the Planck scale. 
	In this work, we are interested in reproducing the effective Friedmann equation of LQC, by considering a generic $f(R,P,Q)$ theory of gravity, where $R=g^{\mu\nu}R_{\mu\nu}$ is the Ricci scalar, $P=R_{\mu\nu}R^{\mu\nu}$, and $Q=R_{\a\b\mu\nu}R^{\a\b\mu\nu}$ is the Kretschmann scalar.
	An order reduction technique allows us to work in $f(R,P,Q)$ theories which are perturbatively close to General Relativity, and to deduce a modified Friedmann equation in the reduced theory.
	Requiring that the modified Friedmann equation mimics the effective Friedmann equation of LQC, we are able to derive several functional forms of $f(R,P,Q)$. We discuss the necessary conditions to obtain viable bouncing cosmologies for the proposed effective actions of $f(R,P,Q)$ theory of gravity.
	\end{abstract}  
	
	\maketitle
	
	%%%%%%%%%%%%%%%%%%%%%%%%%%%%%%%%%%%%%%%%%%%%%%%%%%%%%%%%%%%%%%%%%
	\section{Introduction}
	%%%%%%%%%%%%%%%%%%%%%%%%%%%%%%%%%%%%%%%%%%%%%%%%%%%%%%%%%%%%%%%%%
	
	The cosmological model which received a great consensus from the scientific community is the so-called \textit{Big Bang theory}. Generally, from purely classical considerations, starting from the present day and ideally moving back in time, we arrive at a universe that originated from an initial singularity, namely, the Big Bang singularity, where the volume approaches zero and the energy density and temperature both diverge. Note, however, that one may consider that the singularities present in a theory arise from simplified assumptions and essentially highlight the invalidity of the theory in specific limits and regions. More specifically, the presence of a singularity is a characteristic of the model which one is using to describe a physical system and does not necessarily correspond to an actual physical singularity. Thus, the generalized opinion is that singularities may be cured by taking into account quantum gravity effects, that may become important in extreme scenarios in which gravity dominates over other interactions. Indeed, a quantum theory aimed at solving this singularity is Loop Quantum Cosmology (LQC)\z\cite{Bojowald:2005epg,Bojowald:2012xy,Ashtekar:2011ni,Agullo:2016tjh}. Here, the quantum geometry generates a short-range repulsive force which grows dramatically in the Planck regime (otherwise it is entirely negligible), and which consequently renders the Big Bang singularity into a \textit{quantum bounce}\z\cite{Bojowald:2001xe,Bojowald:2018sgf}.
	
	Rather than discussing all of the fine details of LQC, which is indeed not the aim of this paper, we focus on one of its achievements, namely, the effective Friedmann equation given by\z\cite{Singh:2006sg}:
	\be
	H^2=\frac{1}{3}\kappa \rho\left(1-\frac{\rho}{\rho_c}\right)\,,\label{f1}
	\ee
	where $H=\dot{a}/a$ is the Hubble rate, $a=a(t)$ is the scale factor, the \textit{overdot} denotes a derivative with respect to the cosmological time $t$, $\k=8\pi\,G_{N}/c^4$, with $G_N$ and $c$ being the Newton constant and the speed of light, respectively, $\rho$ is the energy density, and $\rho_c = c^2\sqrt{3}/(32\pi^2\gamma^3 G_N \ell^2_{P})$ corresponds to the critical energy density for the bounce, with $\gamma\approx 0.2375$ and $\ell_{P}=\sqrt{\hslash G_{N}/c^3}$ being respectively the Barbero--Immirzi parameter\z\cite{Lim:2021ney} and the Planck length. Throughout this work, we use  Planck units, where $c=\hslash=G_N=k_B=1$. 
	
	Notice that Eq.\z\eqref{f1} is the standard Friedmann equation of GR with an additive source, which does not in fact increase the original degrees of freedom of the theory. Therefore, the quadratic $\r^2$ term becomes important at very high energy densities while, in counterpart, for $\r\ll\r_c$, the correction $\r/\r_c$ becomes negligible and we recover the standard Friedmann equation. Indeed, it can also be noted that by taking the classical limit $\hslash\rightarrow0$, which implies $\r_c\rightarrow\infty$, and the effective LQC, Friedmann equation reduces to the classical one.	The bounce takes place when $\r=\r_c$, which results in $H^2=0$. 
	In fact, another condition we need to impose in order to have a bounce is that $\ddot{a}/a>0$,  implying  an expansion after the contraction phase (see Refs.\z\cite{Taveras:2008ke,Singh:2006im}). In order to estimate $\ddot{a}/a>0$, taking into account a homogeneous isotropic cosmological perfect fluid in a Friedmann--Lemaître--Robertson--Walker (FLRW) metric, it is necessary to consider the standard energy-momentum conservation law, given by $\dot{\r}+3H(\r+p)=0$, where $p$ is the isotropic pressure.
	Differentiating Eq.\z\eqref{f1} with respect to the cosmological time and using the conservation law, it is straightforward to arrive at an equation for the time derivative for the Hubble rate, $\dot{H}=-\frac{1}{2}\k\,(\r+p)\left(1-2\r/\r_c\,\right)$, which can be rewritten as
	\begin{gather}
		\frac{\ddot{a}}{a}=-\frac{1}{6}\k\,(\r+3p)\left(1-2\,\frac{2\r+3p}{\r+3p}\,\frac{\r}{\r_c}\,\right),
		\label{f2}
	\end{gather}
	which is sometimes denoted as the modified Raychaudhuri equation\z\cite{Singh:2015jus,Marto:2013soa}. It is interesting to notice that by a redefinition of the pressure and energy density in terms of new effective variables\z\cite{Marto:2013soa}, given by
	\begin{gather}
		\r_{\rm{eff}}=\r\left(1-\frac{\r}{\r_c}\right),\qquad p_{\rm{eff}}= p\left(1-2\frac{\r}{\r_c}\right)-\frac{\r^2}{\r_c}\,,
	\end{gather}
	the Friedmann equations recover their standard form,
	\begin{gather}
		H^2=\frac{1}{3}\k\r_{\rm{eff}}\,,\qquad\frac{\ddot{a}}{a}=-\frac{1}{6}\k(\r_{\rm{eff}}+3p_{\rm{eff}})\,,
	\end{gather}
	where $\r_{\rm{eff}}$ and $p_{\rm{eff}}$ also obey the conservation law, $\dot{\r}_{\rm{eff}}+3H(\r_{\rm{eff}}+p_{\rm{eff}})=0\frac{{}_{}}{{}}$.
	
	One usually assumes a barotropic equation of state, $p = w\rho$, for the cosmological perfect fluid, where \textit{a priori} $w=w(\r)$ is not constant ($e.g.$ Ref.\z\cite{Sami:2006wj}), so that the conservation law reduces to $\dot{\r}=-3H(1+w)\r$, which yields the modified Friedmann equations, namely, Eq.\z\eqref{f1} and
	\begin{equation}
	\frac{\ddot{a}}{a}=-\frac{1}{6}\kappa\rho(1+3w)\!\left(1-2\,\frac{2+3w}{1+3w}\,\frac{\rho}{\rho_c}\,\right)\label{ff2},
	\end{equation}
	respectively. The simplest case is for a constant parameter $w$, and this is precisely the situation we consider throughout this work.
	Evaluating Eq.\z\eqref{ff2} at $\r=\c$, results in $\ddot{a}/a=\frac{1}{2}\k\c(w+1)$, and considering that $\ddot{a}/a$ be positive for $\r=\r_c$, then it follows that $w>-1$\z\footnote{For classical perfect fluids, we have $0\leq w\leq1$. However, when dark energy is also taken into account, the range $-1\leq w\leq1$ is assumed (see Refs.\z\cite{Ha:2012ogh,Vikman:2004dc}). Here, we do not want to deal with constraints of $w$ in general but, in this context, it is immediate that $w$ is strictly greater than $-1$.} (see also Ref.\z\cite{Singh:2006im}).
	
	In this paper, we take in account the situation described by the above equations in order to find useful contributions to build up an effective action for gravity such that it reproduces the results obtained by LQC but provides General Relativity (GR) for $\c\to\infty$.
	Specifically, our aim is to adopt a metric \av loop-inspired\cv\z modified gravity, in an isotropic context, in order to reproduce the effective Friedmann equation of LQC. Initially, we consider a general constant equation of state parameter $w$, and finally we will fix $w=1$ which corresponds to a massless scalar field in LQC.	
    To achieve our goal, we follow closely the analysis outlined in Refs.\z\cite{Sotiriou:2008ya,Terrucha:2019jpm,Barros:2019pvc,Bajardi:2020fxh,Ribeiro:2021gds}, which is based on a \textit{reduction technique} of the order of the differential equations\z\cite{Bel:1985zz,Simon:1990ic,Simon:1991bm}. This approach will allow us to obtain solutions which are \textit{perturbatively close} to GR. Namely, first of all, we note that Eq.\z\eqref{f1} can be interpreted as the classical Friedmann equation with a modified source. Then, taking into account that the field equations of metric modified theories of gravity can be written as modified Einstein field equations, namely, $G_{\mu\nu}=\k T^{(m)}_{\mu\nu}+T^{(\rm curv)}_{\mu\nu}$, where $T^{(\rm curv)}_{\mu\nu}$ is an additive energy-momentum tensor due to the higher order curvature terms of the theory (see Refs.\z\cite{Capozziello:2011et, Nojiri:2010wj, Nojiri:2017ncd,Lobo:2008sg,Capozziello:2013vna,Capozziello:2014bqa,Capozziello:2018ddp,CANTATA:2021ktz}). This leads us to connect the LQC corrections of Eq.\z\eqref{f1} to the additional contributions coming from the curvature energy-momentum tensor characterizing modified theories of gravity. The identification is done by using a perturbative approach from GR. Once the perturbation parameter $\eps$ has been introduced, we compare the term $-\k\r^2/3\c$ of Eq.\z\eqref{f1} to the $T^{(\rm curv)}_{00}$ component at the first perturbative order in $\epsilon$\z\footnote{
    This step will be more rigorously explained in the following part of the section.}.

	In particular, in this work, we consider a general $f(R,P,Q)$ theory of gravity, where $R=g^{\mu\nu}R_{\mu\nu}$ is the Ricci scalar, $P=R_{\mu\nu}R^{\mu\nu}$, and $Q=R_{\a\b\mu\nu}R^{\a\b\mu\nu}$ is the Kretschmann, being $ R^{\a}{}_{\b\mu\nu}$ the Riemann tensor and being $R_{\mu\nu}=R^{\lambda}{}_{\mu\lambda\nu}$ the Ricci tensor.
	However, this type of theories is characterized by higher-order derivatives of the metric tensor which generically provide spurious degrees of freedom.
	The order reduction is a way to get around the problem and provides solutions which are perturbatively close to GR.
	The reduction technique consists in a redefinition of $f(R,P,Q)\rightarrow R+\epsilon\,\f(R,P,Q)$, where $\eps$ is the perturbative parameter.
	In general, the dimensionless parameter $\epsilon$ indicates the deviation of the model from GR and, $\varphi(R,P,Q)$ represents a function incorporating all possible corrections to the Einstein--Hilbert action. Thus, $\epsilon$ is designed so that when it approaches to zero GR is restored. Indeed, in our parameterization $f(R,P,Q)=R+\epsilon\,\varphi(R,P,Q)$, if $\epsilon$ is set to zero, the action becomes the standard Einstein--Hilbert one. In this sense, $\epsilon$ indicates the deviations from GR and it can be absorbed in $\varphi(R,P,Q)$. 
    For the order reduction technique to be valid, it is necessary that $\epsilon\varphi \ll R$ at the range of curvatures considered (here, essentially, $R\ll \rho_c \sim l_{p}{}^{-2}$).
    In other words, considering the order $\epsilon$ means that we are working at first order in the correction $\varphi(R,P,Q)$ of the effective action without reaching the critical quantum regime.

	We consider the FLRW metric in order to obtain the first Friedmann equation, then using the GR field equations as zeroth-order perturbative equations, we express our geometric variables, (\textit{i.e}, $R$, $P$ and, $Q$) in terms of energy-matter fields. In this way, we obtain a modified Friedmann equation with an additive term depending on the first order of the perturbation, $H^2=\k\r /3+\eps \,\bar{T}^{(\rm curv)}_{00}/3$, where $\bar{T}^{(\rm curv)}_{00}$ is simply the $0$--$0$ component of $T^{(\rm curv)}_{\mu\nu}$ evaluated at zeroth perturbative order (we will introduce the \av\textit{bar}\cv\z notation to indicate a \av reduced quantity\cv). Consequently, equating $\eps \,\bar{T}^{(\rm curv)}_{00}$ with the LQC correction $-\k\r^2/\c$ of Eq.\z\eqref{f1}, we finally obtain a differential equation, in which the solutions represent possible contributions to the Lagrangian which provide the LQC Friedmann equation\z\eqref{f1}. 
    The reduction technique allows to control efficiently the deviation from GR, and it works in a given curvature regime: we have $R\sim \rho$ while the deviation from GR is $\epsilon\,\varphi\sim \rho^2/\c$. Therefore, the approximation we use is valid for $R\ll\rho_c\sim l_{p}{}^{-2}$. At $\rho\simeq\rho_c$ the approximation breaks down.
    Nevertheless, our position is that, at stages close to the bounce, the effective Lagrangian obtained from the effective Friedmann equation of LQC is still a good approximation describing a collapsing universe just before and just after the bounce.
    Generally, we can say that the approach is valid for $\rho\sim 0.1\rho_c$, 
    which corresponds to a typical length $l\sim 3l_p$.

    However, it is worth noticing that this approach leads to another limitation. Indeed, it is only possible to find some specific (and not all) terms which added to the Einstein--Hilbert action provide the effective Friedmann equation of LQC. Therefore, we cannot build the \av ultimate\cv\z effective $f(R,P,Q)$ theory of gravity due to the fact that we are working with a multi-variable function (see also Refs.\z\cite{Sotiriou:2008ya,Terrucha:2019jpm}).
    
    One of our goals is to generalize the results obtained in the Refs.\z\cite{Sotiriou:2008ya,Terrucha:2019jpm,Barros:2019pvc} which consider $f(R)$, $f(\mathcal{G})$ and $f(R,\mathcal{G})$, respectively, where $\mathcal{G}=Q-4P+R^2$ is the Gauss--Bonnet topological invariant (see Refs.\z\cite{Sotiriou:2008rp,DeFelice:2010aj,Nojiri:2005jg,Bajardi:2020osh,Nojiri:2006ri,Bogdanos:2009tn,Nojiri:2017ncd,Odintsov:2018nch} for more details about these modified theories of gravity).
    
	The paper is organised as follows. In Sec.~\ref{model} we describe the general action for $f(R,P,Q)$ gravity and the respective field equations, and then deploy the order reduction technique. In Sec.~\ref{sectionFLRW}, we focus on a fixed metric tensor and derive the modified Friedmann equation together with its reduced form. In Sec.\z\ref{solutions}, we propose specific functional forms of $f(R,P,Q)$ and analyse in great detail the validity of the parameter range of the functions considered in order to obtain bouncing cosmologies. Finally, we conclude in Sec.~\ref{conclusions}. 
		
	%%%%%%%%%%%%%%%%%%%%%%%%%%%%%%%%%%%%%%%%%%%%%%%%%%%%%%%%%%%%%%%%%
	\section{Fourth-order gravity and order reduction technique}\label{model}
	%%%%%%%%%%%%%%%%%%%%%%%%%%%%%%%%%%%%%%%%%%%%%%%%%%%%%%%%%%%%%%%%%	
	
	Let us consider the following class of gravity theories in a 4-dimensional spacetime,
	\be
	\label{action}
	\mathcal{S}= \frac{1}{2\k}\int d^4 x \sqrt{-g}\, f(R,P,Q)  + \mathcal{S}_m (g_{\mu\nu},\psi)\,,
	\ee
	where the quantities $R$, $P$, and $Q$ are the curvature invariants defined in the Introduction, and $\mathcal{S}_m (g_{\mu\nu},\psi)$ is the matter action, with matter minimally coupled to the metric and $\psi$ collectively denotes the matter fields. Varying the action~\eqref{action} with respect to the metric $g^{\mu\nu}$, yields the following gravitational field equations:
	\bea\label{feq}
	f_{,R}\,G_{\mu\nu}&=&\k T_{\mu\nu}+\frac{1}{2}g_{\mu\nu}f-\frac{1}{2}g_{\mu\nu}{f}_{,R}R-\nabla_{\mu}\nabla_{\nu}{f}_{,R}+g_{\mu\nu}\Box {f}_{,R}-2f_{,Q}R_{\a\b\g\mu}R^{\a\b\g}{}_{\nu}+4\nabla_{\a}\nabla_{\b}[f_{,Q}R^{\a}{}_{(\mu\nu)}{}^{\b}]\nn\\&&-2{f}_{,P}R^{\a}{}_{\mu}R_{\a\nu}+2\nabla_{\a}\nabla_{\b}[f_{,P}R^{\a}{}_{(\mu}\delta^{\b}{}_{\nu)}]-\Box[f_{,P}R_{\mu\nu}]-g_{\mu\nu}\nabla_{\a}\nabla_{\b}[f_{,P}R^{\a\b}]\,,
	\eea
	where $\nabla_{\a}$ denotes the covariant derivative associated to the Levi--Civita connection, $\Box=g^{\a\b}\nabla_\a\nabla_\b$ is the d'Alembert operator, the comma denotes a partial derivative, \textit{i.e.}, $f_{,R}=\p{f}/\p{R}$, while the parenthesis in the subscript indicates the symmetric part of a tensor with respect to the indices inside them, \textit{i.e.}, $S_{(ab)c}=(S_{abc}+S_{bac})/2$. The energy-momentum tensor of the matter $T_{\mu\nu}$ is defined in the standard manner, namely,
	\be
	T_{\mu\nu}=-\frac{2}{\sqrt{-g}} \frac{\delta S_{m}}{\delta g^{\mu\nu}}\,.  \ee

	It is worth noticing that important subcases as $f(R)$, $f({\cal G})$, and $f(R, {\cal G})$ are included in $f(R,P,Q)$ theories of gravity. 
	
	Modified theories of gravity, such as the one we are going to analyse in this work, are characterized by the presence of further degrees of freedom. However, a way to restore the original degrees of freedom related to GR is by considering an order reduction, which essentially consists in a parameterization of $f(R,P,Q)$ as
	\be\label{phi}
	f(R,P,Q)=R+2\L+\eps\,\f(R,P,Q)\,,
	\ee
	where $\L$ is a cosmological constant term and $\eps$ is a dimensionless perturbative parameter. This parameterization is such that, substituting Eq.\z\eqref{phi} into Eqs.\z\eqref{action}\z and\z\eqref{feq}, for $\eps=0$ provides GR. Thus, it allows us to interpret $f(R,P,Q)$ gravity as an effective field theory with solutions perturbatively close to GR, when $\eps\f\ll R$. The order reduction technique is then implemented by evaluating Eq.\z\eqref{feq} to the first order in $\eps$. However, this is not enough due to the presence of the Riemann tensor, and one needs a further simplification. To this effect, we assume that our spacetime is conformally flat, \textit{i.e.}, there exists always a local reference frame where the metric is flat (Minkowski metric) up to a conformal factor. The reason for this assumption is that we are interested in studying this modified theory of gravity from a cosmological point of view and, in particular, we will work with the FLRW metric which always fulfils this property. In a conformally flat spacetime, the Weyl tensor $C_{\alpha\beta\mu\nu}$, which is the traceless part of the Riemann tensor, vanishes identically. Thus, it is possible to express the Riemann tensor in terms of the Ricci tensor and the Ricci scalar only, as follows:
	\be\label{riemann}
	R_{\alpha\beta\mu\nu} \,=   \frac{1}{2}\left( g_{\mu\alpha}R_{\beta\nu}+g_{\nu\beta}R_{\alpha\mu}- g_{\mu\beta}R_{\alpha\nu}  -g_{\nu\alpha}R_{\beta\mu} \right) +\frac{1}{6}\left( g_{\mu\beta}g_{\alpha\nu} - g_{\mu\alpha}g_{\beta\nu} \right) R\,.
	\ee
    As mentioned in the Introduction, we use a \av \emph{bar}\cv\z to indicate a quantity evaluated at the zeroth-order with respect to $\eps\,$ (the reduced quantity). Then, Eq.\z\eqref{feq} at the zeroth-order reduces to
    \begin{equation}\label{GR}
    \bar{R}_{\mu\nu}-\frac{1}{2}g_{\mu\nu}\bar{R}=\k T_{\mu\nu}-g_{\mu\nu}\L\,.
    \end{equation}
	Therefore, we get the expression of $R$ at the zeroth-order with respect to $\eps$ considering the trace of Eq.\z\eqref{GR}, given by
	\begin{equation}
		\bar{R} = \,\,-4\Lambda-\kappa T \,, \label{rR}
	\end{equation}
	 and we obtain the reduced forms of $R_{\mu\nu}$ and $R_{\a\b\mu\nu}$, by using Eqs.\z\eqref{riemann}\z--\z\eqref{rR}, provided by 
	\bea
		\bar{R}_{\mu\nu} &=& - \Lambda g_{\mu\nu}+\kappa T_{\mu\nu}-\frac{\kappa}{2}g_{\mu\nu}T \,,\label{ricci}\\
		\bar{R}_{\mu\nu\alpha\beta} &=&-\frac{\kappa}{2}\left( g_{\mu\beta}T_{\alpha\nu}+g_{\nu\alpha}T_{\beta\mu} - g_{\mu\alpha}T_{\beta\nu}-g_{\nu\beta}T_{\alpha\mu} \right) -\frac{1}{3}\left( g_{\mu\alpha}g_{\nu\beta}-g_{\mu\beta}g_{\alpha\nu}\right) \left( \Lambda+\kappa T \right)\,, \label{riem}
	\eea
	respectively,
	where $T=g^{\mu\nu}T_{\mu\nu}$ is the trace of the energy-momentum tensor.
	
	The last two quantities we need are the reduced form of our scalar variables, namely,
	\bea
		&\bar{P}&\,=4\L^2+2\k T\L+\k^{2}T^{\mu\nu}T_{\mu\nu}\,\label{rP},\\
		&\bar{Q}&\,=\frac{8}{3}\L^2+\frac{4}{3}\k T\L-\frac{1}{3}\k^2 T^2+2\k^2 T_{\mu\nu}T^{\mu\nu}\,.\label{rQ}
	\eea

	At this point, we have all the ingredients to obtain the order reduced field equations, at first order in $\eps$, by substituting Eqs.\z\eqref{phi}\z--\z\eqref{rQ} in Eq.\z\eqref{feq}, which finally yields:
	\bea\label{rfeq}
	G_{\mu\nu}&=&\k T_{\mu\nu}+\L g_{\mu\nu} +\eps\,\bigg[\frac{1}{2}g_{\mu\nu}\bar{\f}-\bar{R}_{\mu\nu}\bar{\f}_{,\bar{R}}-\nabla_{\mu}\nabla_{\nu}{\bar{\f}}_{,\bar{R}}+g_{\mu\nu}\Box {\bar{\f}}_{,\bar{R}}-2\bar{\f}_{,\bar{Q}}\bar{R}_{\a\b\g\mu}\bar{R}^{\a\b\g}{}_{\nu}+4\nabla_{\a}\nabla_{\b}(\bar{\f}_{,\bar{Q}}\bar{R}^{\a}{}_{(\mu\nu)}{}^{\b})
	\nn\\
	&& -2{\bar{\f}}_{,\bar{P}}\bar{R}^{\a}{}_{\mu}\bar{R}_{\a\nu}+2\nabla_{\a}\nabla_{\b}(\bar{\f}_{,\bar{P}}\bar{R}^{\a}{}_{(\mu}\delta^{\b}{}_{\nu)})-\Box(\bar{\f}_{,\bar{P}}\bar{R}_{\mu\nu})-g_{\mu\nu}\nabla_{\a}\nabla_{\b}(\bar{\f}_{,\bar{P}}\bar{R}^{\a\b})\,\bigg],
	\eea
	where $\bar{\f}=\f(\bar{R},\bar{P},\bar{Q})$. These are the field equations that will be used throughout this work, in order to obtain the effective Friedmann equation\z\eqref{f1} of LQC in our setting.

	%%%%%%%%%%%%%%%%%%%%%%%%%%%%%%%%%%%%%%%%%%%%%%%%%%%%%%%%%%%%%%%%%
	\section{Modified Friedmann equation and bouncing cosmology}\label{sectionFLRW}
	%%%%%%%%%%%%%%%%%%%%%%%%%%%%%%%%%%%%%%%%%%%%%%%%%%%%%%%%%%%%%%%%%
	
	In order to derive Eq.\z\eqref{f1}, let us consider the FLRW line element:
	\be
	ds^2 = -dt^2 +a(t)^2 \left[ \frac{dr^2}{1-kr^2} +r^2\left( d\theta^2 + \sin^2\theta\, d\phi^2 \right) \right],
	\ee
	where $k=-1,0,1$ corresponds to a hyperbolic, flat, and hyperspherical spatial curvature, respectively. Furthermore, consider an energy-momentum tensor describing a homogeneous isotropic cosmological perfect fluid, \textit{i.e.}, $T_{\mu\nu} = \left( \rho + p \right)u_{\mu}u_{\nu} + p\,g_{\mu\nu}$,
	where $u_{\mu}$ is the 4-velocity of the fluid, with normalization $u_{\mu} u^{\mu}=-1$, and $p$ and $\rho$ are the isotropic pressure and energy density, respectively. Moreover, as mentioned in the Introduction, we assume a barotropic equation of state $p = w\rho$, with a constant parameter $w>-1$, which corresponds to an accelerated expanding universe in LQC, where the energy-momentum conservation law, $\nabla_{\mu}T^{\mu\nu}=0$, provides the following continuity equation
	\be\label{continuity}
	\dot{\rho}= -3H \left( 1+w \right)\rho\,.
	\ee
	Then, Eqs.\z\eqref{rR},\z\eqref{rP}, and\z\eqref{rQ}, can be rewritten in the following explicit form, respectively,
	\bea
	\bar{R}&=&-4\L+\k(1-3w)\r\label{R}\,,\\
	\bar{P}&=& 4 \Lambda ^2+2 \kappa  \Lambda    (3 w -1)\rho+\kappa ^2  (3 w ^2+1)\rho ^2\,,\label{P}\\
	\bar{Q}&=&\frac{8}{3}\Lambda ^2+\frac{4}{3} \kappa  \Lambda  (3 w -1) \rho+\frac{1}{3} \kappa ^2 (9 w ^2+6 w +5) \rho ^2 \,,\label{Q}
	\eea
	and by using Eq.\z\eqref{continuity}, we obtain their derivatives with respect to cosmological time, given by
	\bea
	\dot{\bar{R}}&=&3 \kappa  (w +1) (3 w -1) H \rho \label{R'}\,,\\
	\dot{\bar{P}}&=& 6 \kappa \Lambda (w +1)(1-3 w ) H \rho    -6 \kappa^2  (w +1)(1+3w ^2) H\rho^2 \,\label{P'},\\
	\dot{\bar{Q}}&=&4 \kappa \Lambda (w +1)(1-3 w )  H \rho   -2 \kappa ^2 (w +1)(9 w ^2+6 w +5)  H \rho^2\,.\label{Q'}
	\eea
	These are useful for computing the time derivative of $\bar{\f}$ (by a change of variables) and to obtain the final form of the field equations.
	
	The first Friedmann equation can be derived by considering the $0\,$--$\,0$ component of Eq.\z\eqref{feq}, and the second Friedmann equation can be obtained by evaluating one spatial component of Eq.\z\eqref{feq}. However, for the sake of brevity, only the first Friedmann equation is given below. Indeed, the second equation, as we have mentioned above, depends on the continuity equation and the first Friedmann equation, and therefore, in the present analysis, it is practically irrelevant.
	
	Taking into account what has been discussed so far, it is possible to see that the reduced modified Friedmann equation reads as:
	\bea
	H^{2}=&&\frac{1}{3}\kappa  \rho -\frac{k}{a^2}-\frac{\Lambda }{3}\nn\\
	&&+\epsilon \, \bigg\{-\frac{1}{6}\bar{\f}+\frac{1}{9}\bigg[{4 \Lambda ^2}+2 (9 w^2+12 w-1) \kappa  \rho  \Lambda + (-9 w^2-6 w+7) \kappa ^2 \rho ^2+18  (w+1) (3 w-1) \frac{k}{a^2}\kappa  \rho\bigg] \bar{\f}_{,\bar{Q}}\nn\\
	&&+\k\r(w+1)\bigg[-\frac{4}{9}   (3 w+1) (9 w^2+6 w+5) \kappa ^3 \rho ^3+\frac{4}{3} { (3 w+1) (9 w^2+6 w+5)\frac{k}{a^2} \kappa ^2 \rho ^2}+\frac{16}{3}  (3 w^2+1)   \Lambda ^2 \k \rho\nn\\
	&&+\frac{4}{3}  (3 w-1) (3 w^2+1) \L \kappa ^2   \rho ^2 +\frac{16}{3} {  (9 w^2+3 w+2)\frac{k}{a^2} \L \kappa    \rho }+\frac{16}{9} (3 w-1)   \Lambda ^3  +\frac{16}{3} { (3 w-1) \frac{k}{a^2}  \Lambda ^2  }\bigg] \bar{\f}_{,\bar{Q},\bar{Q}}\nn\\
	&&+\frac{1}{3}\bigg[{2 \Lambda ^2}+ (6 w^2+9 w-1) \kappa  \rho  \Lambda + (-3 w^2-3 w+2) \kappa ^2 \rho ^2+\frac{k}{a^2} 6(w+1) (3 w-1) \kappa  \rho \bigg] \bar{\f}_{,\bar{P}}\nn\\
	&&+\k\r(w+1)\bigg[-\frac{4}{3} (18 w^3+9 w^2+8 w+1) \kappa ^3 \rho ^3+ {4(18 w^3+9 w^2+8 w+1) \frac{k}{a^2}\kappa ^2 \rho ^2} +8 (15 w^2+2 w+3) \frac{k}{a^2}\L\kappa    \rho \nn\\
	&&+\frac{4}{3}  (18 w^3-21 w^2+4 w-5) \L\kappa ^2  \rho ^2+\frac{8}{3}  (15 w^2-4 w+5)\L^2 \kappa \rho+\frac{16}{3}  (3 w-1)   \Lambda ^3  + 16 (3 w-1)\frac{k}{a^2}   \Lambda ^2  \bigg] \bar{\f}_{,\bar{P},\bar{Q}}\nn\\
	&&+4\k\r(w+1)\bigg[- w  (3 w^2+1) \kappa ^3 \rho ^3+ 3w  (3 w^2+1)\frac{k}{a^2} \kappa ^2 \rho ^2+  (3 w^3-6 w^2+2 w-1) \kappa ^2 \Lambda  \rho ^2\nn\\
	&&+2 (3 w^2-2 w+1) \kappa  \Lambda ^2 \rho +4   (6 w^2-w+1) \frac{k}{a^2}\L\kappa \rho + (3 w-1)   \Lambda ^3  +3  (3 w-1) \frac{k}{a^2}  \Lambda ^2\bigg] \bar{\f}_{,\bar{P},\bar{P}}\nn\\
	&&+\bigg[-\frac{1}{6}(3w+1) \kappa  \rho -\frac{\Lambda }{3}\bigg] \bar{\f}_{,\bar{R}}+\frac{4}{3}\k\r(w+1)\bigg[(9 w^2+3 w+2) \kappa ^2 \rho ^2-(9 w^2-3 w+4) \Lambda\kappa   \rho\nn\\
	&&- 3(9 w^2+3 w+2)\frac{k}{a^2} \kappa  \rho -2 (3 w-1)  \Lambda ^2  -6 (3 w-1) \frac{k}{a^2}\Lambda \bigg] \bar{\f}_{,\bar{R},\bar{Q}}\nn\\
	&&+2\k\r(w+1) \bigg[ (6 w^2-w+1) \kappa ^2 \rho ^2- (6 w^2-7 w+3) \L\kappa   \rho -3 (6 w^2-w+1) \frac{k}{a^2}\kappa \rho-2  (3 w-1)   \Lambda ^2\kappa \rho \nn\\
	&&-6  (3 w-1) \frac{k}{a^2} \Lambda\bigg] \bar{\f}_{,\bar{R},\bar{P}}-\k\r(3 w-1)(w+1)
	\left( \kappa  \rho-3 \frac{k}{a^2}- \Lambda  \right) \bar{\f}_{,\bar{R},\bar{R}}\,\bigg\}\,,
	\label{modFF}
	\eea
	where it was necessary to substitute the zeroth-order expressions of $H^2$ and $\ddot{a}/a$ which, in general, turn out to be
	\begin{equation}
		H^2= \frac{1}{3} \kappa  \rho -\frac{k}{a^2}-\frac{\Lambda }{3} \,, \qquad \frac{\ddot{a}}{a}= -\frac{1}{2}\left( w  +\frac{1}{3}\right) \kappa\rho-\frac{\Lambda }{3}\,.
	\end{equation}
	Thus, following Refs.\z\cite{Sotiriou:2008ya,Terrucha:2019jpm,Barros:2019pvc}, we assume spatial flatness of the spacetime ($k=0$), and noticing that $\L$ does not contribute to Eq.\z\eqref{f1}, we set $\L=0$. Thus, the reduced variables are given by
	\bea
		\bar{R}&=&(1-3w)\k\r\label{Rw}\,,\\
		\bar{P}&=& (3 w ^2+1)\kappa ^2  \rho ^2\,,\label{Pw}\\
		\bar{Q}&=&\frac{1}{3}(9 w ^2+6 w +5)  \kappa ^2 \rho ^2 \,,\label{Qw}
	\eea
	while, the first Friedmann equation\z\eqref{modFF} reduces to
	\bea\label{FF}
	H^2&=&\frac{1}{3} \kappa  \rho+\epsilon\left[-\frac{1}{6}\bar{\f}+\frac{1}{6}  (-3 w-1) \kappa\rho\, \bar{\f}_{,\bar{R}}+\frac{1}{3}  \left(-3 w^2-3 w+2\right) \kappa^2\rho^2 \bar{\f}_{,\bar{P}}+\frac{1}{9}  \left(-9 w^2-6 w+7\right) \kappa^2\rho^2\, \bar{\f}_{,\bar{Q}}\right.\nn\\
	&&- (w+1) (3 w-1) \kappa^2\rho^2\, \bar{\f}_{,\bar{R},\bar{R}}-4 w (w+1) \left(3 w^2+1\right)  \kappa ^4\rho^4\, \bar{\f}_{,\bar{P},\bar{P}}\nn\\
	&&-\frac{4}{9} \left(27 w^4+54 w^3+48 w^2+26 w+5\right) \kappa^4 \rho^4\,\bar{\f}_{,\bar{Q},\bar{Q}}+2  (w+1) \left(6 w^2-w+1\right) \kappa^3\rho^3\,\bar{\f}_{,\bar{\bar{R}},\bar{P}}\nn\\
	&&\left.+\frac{4}{3}  (w+1) \left(9 w^2+3 w+2\right) \kappa^3\rho^3\,\bar{\f}_{,\bar{R},\bar{Q}}-\frac{4}{3}  (w+1) \left(18 w^3+9 w^2+8 w+1\right) \kappa^4\rho^4\,\bar{\f}_{,\bar{P},\bar{Q}}\right].
	\eea
	Then, comparing the above modified Friedmann equation with Eq.\z\eqref{f1}, it turns out that $\f$ has to satisfy the following differential equation,
	\bea\label{de_FF}
	&&-\frac{1}{6}\bar{\f}+\frac{1}{6}  (-3 w-1) \kappa\rho\, \bar{\f}_{,\bar{R}}+\frac{1}{3}  \left(-3 w^2-3 w+2\right) \kappa^2\rho^2 \bar{\f}_{,\bar{P}}+\frac{1}{9}  \left(-9 w^2-6 w+7\right) \kappa^2\rho^2\, \bar{\f}_{,\bar{Q}}\nn\\
	&&- (w+1) (3 w-1) \kappa^2\rho^2\, \bar{\f}_{,\bar{R},\bar{R}}-4 w (w+1) \left(3 w^2+1\right)  \kappa ^4\rho^4\, \bar{\f}_{,\bar{P},\bar{P}}\nn\\
	&&-\frac{4}{9} \left(27 w^4+54 w^3+48 w^2+26 w+5\right) \kappa^4 \rho^4\,\bar{\f}_{,\bar{Q},\bar{Q}}+2  (w+1) \left(6 w^2-w+1\right) \kappa^3\rho^3\,\bar{\f}_{,\bar{\bar{R}},\bar{P}}\nn\\
	&&+\frac{4}{3}  (w+1) \left(9 w^2+3 w+2\right) \kappa^3\rho^3\,\bar{\f}_{,\bar{R},\bar{Q}}-\frac{4}{3}  (w+1) \left(18 w^3+9 w^2+8 w+1\right) \kappa^4\rho^4\,\bar{\f}_{,\bar{P},\bar{Q}}=-\frac{\k^2\r^2}{3\eps\k\c}\,.
	\eea
	In particular, fixing $w=1$, the first Friedmann equation simplifies further to 
	\bea
	H^{2}&=&\frac{1}{3}  \kappa  \rho +\epsilon  \left(-\frac{1}{6}\bar{\f}\right.-\frac{2}{3} \kappa  \rho \, \bar{\f}_{,\bar{R}}-\frac{4}{3} \kappa ^2 \rho ^2 \bar{\f}_{,\bar{P}}-\frac{8}{9} \kappa ^2 \rho ^2 \bar{\f}_{,\bar{Q}}+24 \kappa ^3 \rho ^3 \bar{\f}_{,\bar{R},\bar{P}}-96 \kappa ^4 \rho ^4 \bar{\f}_{,\bar{P},\bar{Q}}+\frac{112}{3} \kappa ^3 \rho ^3 \bar{\f}_{,\bar{R},\bar{Q}}\nn\\
	&&\left.-4 \kappa ^2 \rho ^2 \bar{\f}_{,\bar{R},\bar{R}}-32 \kappa ^4 \rho ^4 \bar{\f}_{,\bar{P},\bar{P}}-\frac{640}{9} \kappa ^4 \rho ^4 \bar{\f}_{,\bar{Q},\bar{Q}}\right),\label{w=1}
	\eea
	and the associated differential equation is given by
	\bea
	 &&-\frac{1}{6}\bar{\f}-\frac{2}{3} \kappa  \rho \, \bar{\f}_{,\bar{R}}-\frac{4}{3} \kappa ^2 \rho ^2 \bar{\f}_{,\bar{P}}-\frac{8}{9} \kappa ^2 \rho ^2 \bar{\f}_{,\bar{Q}}+24 \kappa ^3 \rho ^3 \bar{\f}_{,\bar{R},\bar{P}}-96 \kappa ^4 \rho ^4 \bar{\f}_{,\bar{P},\bar{Q}}\nn\\
	&&+\frac{112}{3} \kappa ^3 \rho ^3 \bar{\f}_{,\bar{R},\bar{Q}}-4 \kappa ^2 \rho ^2 \bar{\f}_{,\bar{R},\bar{R}}-32 \kappa ^4 \rho ^4 \bar{\f}_{,\bar{P},\bar{P}}-\frac{640}{9} \kappa ^4 \rho ^4 \bar{\f}_{,\bar{Q},\bar{Q}}=-\frac{\k^2\r^2}{3\eps\k\c}\,.\label{de_w=1}
	\eea
	Equations\z\eqref{de_FF} and\z\eqref{de_w=1} are non-linear second-order partial differential equations for a three-variable function. Therefore, it is not possible to determine the general form of $\f$ such that Eqs.\z\eqref{FF} and\z\eqref{w=1} be equal to Eq.\z\eqref{f1}.
	Thus, we need to propose some specific \textit{ansatz} on $f(R,P,Q)$ and $\f$, and investigate the conditions under which the action\z\eqref{action} can reproduce a bouncing universe. In this regard, in order to construct a \av loop-inspired modified gravity\cv\z\footnote{Here we are considering Eq.\z\eqref{f1} without further corrections as in Ref.\z\cite{Li:2019ipm}. Therefore, we are not entitled to consider different cases than $w=1$, if we want to have a strict consistency with LQC.}, notice that $\bar{R}$ is the only scalar among our variables which can be positive, negative, or equal to zero, while $\bar{P}$ and $\bar{Q}$ are strictly positive, which is transparent from Eqs.\z\eqref{Rw}\z--\z\eqref{Qw}. Therefore, we should pay special attention to the case in which $\bar{R}=0$, $i.e.$, $w=1/3$. Another particular case is $w=-1/3$ which corresponds to $\bar{\mathcal{G}}=0$, where $\bar{\mathcal{G}}=\frac{2}{3}\bar{R}^2-2\bar{P}=-\frac{4}{3}(1+3w)k^2\r^2$ is the reduced Gauss--Bonnet term. Finally, recall that we consider only values of $w>-1$, due to the condition of $\ddot{a}/a>0$ for $\r=\c$.
	
	%%%%%%%%%%%%%%%%%%%%%%%%%%%%%%%%%%%%%%%%%%%%%%%%%%%%%%%%%%%%%%%%%	
	\section{Specific functional forms of $f(R,P,Q)$ theory of gravity}\label{solutions}
	%%%%%%%%%%%%%%%%%%%%%%%%%%%%%%%%%%%%%%%%%%%%%%%%%%%%%%%%%%%%%%%%%	
	
	%%%%%%%%%%%%%%%%%%%%%%%%%%%%%%%%%%%%%%%%%%%%%%%%%%%%%%%%%%%%%%%%%	
	\subsection{Solution I}\label{solution1}
	%%%%%%%%%%%%%%%%%%%%%%%%%%%%%%%%%%%%%%%%%%%%%%%%%%%%%%%%%%%%%%%%%	
	
	First of all, we aim to generalize the results of Ref.\z\cite{Sotiriou:2008ya} and Ref.\z\cite{Terrucha:2019jpm}, which discuss effective $f(R)$ and $f(\mathcal{G})$ bouncing cosmologies, assuming $f(R)=R+\eps\f(R)$ and $f(\mathcal{G})=R+\eps\f(\mathcal{G})$, respectively\z\footnote{Actually, in both treatments there is also the presence of the cosmological constant $\L$ inside the gravitational action. However, they set $\L=0$ for our same reason.}.
	In Refs.\z\cite{Sotiriou:2008ya,Terrucha:2019jpm}, the perturbative function $\f$ depends only on a single variable, while in the present work we are essentially dealing with three variables. 
	
	In order to simplify our approach, and to arrive at a situation similar to $f(R)$ and $f(\mathcal{G})$, we need to find a way to work with a single variable, which is a generalization of $R$ and $\mathcal{G}$. To this effect, we define a new variable $\mathcal{X}=\mathcal{X}(R,P,Q)$, and \av perturb\cv\z the GR Lagrangian by using a function of $\mathcal{X}$. In particular, we assume that
	\be
	f(R,P,Q)=R+\eps\f\left(\mathcal{X}\right),\qquad{\rm with} \qquad \enspace\mathcal{X}=C_1\abs{R}^{2\a}+C_2\,P^\a+C_3\,Q^\a\label{ansatz1}\,,
	\ee
	where $\a$ is a dimensionless real parameter, and the constants $\{C_i\}$ with $i=1,2,3$ possess the correct dimensions such that the variable $\mathcal{X}$ is dimensionless\z\footnote{In our case, the following dimensional equation holds: $[C_i]=[\k\r]^{-2\a}=[\r]^{-2\a}$, with $i=1,2,3$.}. The variable $\mathcal{X}$ is the most general linear combination of the powers of $R$, $P$ and $Q$ we can consider in this context. The reason is that in order to give a well posed definition of $\mathcal{X}$, all the quantities in Eq.\z\eqref{ansatz1} must have homogeneous dimensions. Moreover, to apply the reduction method of Refs.\z\cite{Sotiriou:2008ya,Terrucha:2019jpm}, we need an invertible relation between $\mathcal{X}$ and the energy-matter density $\rho$.
	In addition, notice that $R$ can assume positive and negative values, depending on the value of $w$, while $P$ and $Q$ are always positive defined, as mentioned above. Therefore, the absolute value of $R$ allows us to consider exponents that take on real values.
	Finally, it is easy to see that: for $\a=1/2,\,C_1=1,\,C_2=0=C_3$, we get $\mathcal{X}=R$; for $\a=1,\,C_1=1,\,C_2=-4,\,C_3=1$, we get $\mathcal{X}=\mathcal{G}$.

	At this point, we want to use the assumption\z\eqref{ansatz1} to solve Eq.\z\eqref{de_FF}. Therefore, we need to write the reduced form of $\mathcal{X}$. Using Eqs.\z\eqref{Rw}\z--\z\eqref{Qw}, it is straightforward to see that $\bar{\mathcal{X}}$ is given by
    \be
    \bar{\mathcal{X}}= C\, (\kappa  \rho )^{2 \alpha }\,,\quad{\rm where}\quad C= C_1 \left| 1-3 w  \right| ^{2 \alpha }+C_2 \left(3 w  ^2+1\right)^{\alpha }+3^{-\alpha }\,C_3 \left(9+6w+5\right)^{\alpha }\,.\label{redX}
    \ee
	We are assuming to work with all possible configurations of parameters $\{w>-1,\a,C_i\}$ that correspond to a non-vanishing variable $\bar{\mathcal{X}}$, \textit{i.e.}, $C\neq0$. Moreover, let us notice that the case $w=1/3$ (corresponding to $R=0$) needs to be analysed separately. Finally, we point out that, from the Eq.\z\eqref{redX}, for fixed values of constant parameters we have ${\rm sign}(\bar{\mathcal{X}})={\rm sign}(C)$ where $C$ is constant. Therefore, $\bar{\mathcal{X}}$ will be either strictly positive definite or strictly negative definite. 
	
	We can now obtain the reduced Friedmann equation\z\eqref{FF} by making a change of variable, from $R,\,P,\,Q$ to $\mathcal{X}=\mathcal{X}(R,P,Q)$. Thus, we obtain the following equation
	\begin{align}
		H^{2}=\frac{1}{3} \kappa  \rho+\epsilon\,\bigg\{-\frac{1}{6} {\f}(\bar{\mathcal{X}})+A\, {\f}'(\bar{\mathcal{X}}) (\kappa \rho)^{2\a}+ B\, {\f}''(\bar{\mathcal{X}})(\kappa\rho)^{4\a}\,\bigg\}\,,
		\label{F_ans1}
	\end{align}
	where the prime corresponds to the derivative with respect to the variable $\bar{\mathcal{X}}$, while $A$ and $B$ are coefficients defined as
	\begin{eqnarray}
		A &=&\left[3\alpha  \left(C_2 (2-3 (4 \alpha -3) w   (w  +1)) \left(3 w  ^2+1\right)^{\alpha -1}+3^{-\alpha } C_3 \left(9 w  ^2+6 w  +5\right)^{\alpha -1} (-12 \alpha  (w  +1) (3 w  +1)\right.\right.
		\nn\\
		&&+\left.\left.3 w   (9 w  +14)+19)-\frac{C_1 \left| 1-3 w  \right| ^{2 \alpha } (12 \alpha  (w  +1)-9 w  -7)}{3 w  -1}\right)\right] \times
		\nn\\
		&& \times \enspace 3^{-\alpha }\left[C_2 \left(9 w  ^2+3\right)^{\alpha }+3^{\alpha } C_1 \left| 1-3 w  \right| ^{2 \alpha }+C_3 \left((3 w  +1)^2+4\right)^{\alpha }\right]^{-1}\,,\\
		B &=&\left[4 \alpha ^2 (w  +1) \left((3 w  -1) \left(C_2^2 w   \left(9 w  ^2+6 w  +5\right) \left(9 w  ^2+3\right)^{2 \alpha }+C_3^2 \left(9 w  ^3+3 w  ^2+3 w  +1\right) \left(9 w  ^2+6 w  +5\right)^{2 \alpha }\right.\right.\right.\nn\\
		&&+\left.\left.\left.3^{\alpha } C_2 C_3 \left(18 w  ^3+9 w  ^2+8 w  +1\right) \left(27 w  ^4+18 w  ^3+24 w  ^2+6 w  +5\right)^{\alpha }\right)\right.\right.\nn\\
		&&+\left.\left.3^{\alpha } C_1 \left(C_2 \left(54 w  ^4+27 w  ^3+33 w  ^2+w  +5\right) \left(9 w  ^2+3\right)^{\alpha }+2 C_3 \left(27 w  ^4+9 w  ^3+15 w  ^2+3 w  +2\right) \times \right.\right.\right.
		\nn\\
		&&\times\left.\left.\left. 
		\left(9 w  ^2+6 w  +5\right)^{\alpha }\right)		
		\left| 1-3 w  \right| ^{2 \alpha }+9^{\alpha } C_1^2 \left(27 w  ^4+18 w  ^3+24 w  ^2+6 w  +5\right) \left| 1-3 w  \right| ^{4 \alpha }\right)\right]  \times
		\nn\\
		&& \times  \left[(3 w  -1) \left(3 w  ^2+1\right) \left(9 w  ^2+6 w  +5\right) \left(3^{\alpha } C_1 \left| 1-3 w  \right| ^{2 \alpha }+C_2 \left(9 w  ^2+3\right)^{\alpha }+C_3 \left((3 w  +1)^2+4\right)^{\alpha }\right){}^2\right]^{-1}.
	\end{eqnarray}

	Then, following an analogous procedure as done in Refs.\z\cite{Sotiriou:2008ya,Terrucha:2019jpm,Barros:2019pvc}, we compare Eq.\z\eqref{F_ans1} with Eq.\z\eqref{f1}, in order to obtain a differential equation for $\f({\bar{\mathcal{X}}})$, \textit{i.e.}, a new version of Eq.\z\eqref{de_FF} with the variable $\bar{\mathcal{X}}$. However, as in Eq.\z\eqref{de_FF}, we have the presence of $\r$ as well as $\bar{\mathcal{X}}$. Thus, using Eq.\z\eqref{redX}, we can rewrite $\r$ as a function of $\bar{\mathcal{X}}$. In this way we obtain a second order differential equation for ${\f}(\bar{\mathcal{X}})$ which can be written in the following form:
	\begin{equation}\label{ECeq}
		-\frac{1}{6} {\f}(\bar{\mathcal{X}})+\frac{A}{C}\,\bar{\mathcal{X}} {\f}'(\bar{\mathcal{X}}) + \frac{B}{C^2}\, \bar{\mathcal{X}}^2 {\f}''(\bar{\mathcal{X}})=-\frac{\left| \bar{\mathcal{X}}\right|^{\frac{1}{\a}}}{3\left|C\right|^{\frac{1}{\a}}\eps\k\c} \,.
	\end{equation}	
	The above differential equation is a second order non-homogeneous Euler--Cauchy equation. This is highly significant because we already know all possible homogeneous and particular solutions, without entering too much into the possible values of the parameters $\{w,\a,C_i\}$.
	
	Before proceeding into the analysis of Eq.\z\eqref{ECeq}, we stress that by choosing $\{\a,\,C_i\}$ such that $\mathcal{X}=R$ or $\mathcal{X}=\mathcal{G}$, Eq.\z\eqref{ECeq} turns out to be the differential equation of Ref.\z\cite{Sotiriou:2008ya} or the one of Ref.\z\cite{Terrucha:2019jpm}. Indeed both differential equations of Ref.\z\cite{Sotiriou:2008ya} and Ref.\z\cite{Terrucha:2019jpm} are a second order non-homogeneous Euler--Cauchy equation. As we will see, the different solutions depend on the coefficients of the differential equation.
	
	In this regard, let us rewrite Eq.\z\eqref{ECeq} in a more appropriate and useful form, defining $x=\left| \bar{\mathcal{X}}\right|$ (dimensionless positive defined variable) and $y(x)={\f}(\bar{\mathcal{X}})\,$:
	\begin{equation}\label{ECx}
		-\frac{1}{6} y(x)+\frac{A}{C}\,x y'(x) + \frac{B}{C^2}\, x^2 y''(x)=-\frac{x^{\frac{1}{\a}}}{3\left|C\right|^{\frac{1}{\a}}\eps\k\c} \,.
	\end{equation}

	At this point, we can distinguish two general nontrivial cases depending on the coefficient of the second order term: $B\neq0$ and $B=0$.\\
  
	Assuming $B\neq0$ and multiplying by $C^2/B$, the differential equation \eqref{ECx} reads as
	\begin{equation}\label{EC}
		x^2y''(x)+axy'(x)+by(x)= cx^{1/\a}\,,
	\end{equation}
	where
	\begin{equation}
		a =\frac{C A}{B}\,,  \qquad b = -\frac{C^2}{6B}\,, \qquad  c =-\frac{1}{3B\left|C\right|^{\frac{1}{\a}-2}\eps\k\c}\,.
	\end{equation}
	In this case, the classification of the homogeneous solutions is based on the roots of the characteristic equation associated to Eq.\z\eqref{EC},
	\begin{equation}\label{characteristic_eq}
		\l^2+(a-1)\l+b=0\,,
	\end{equation}
    which is obtained by substituting the trial solution $y_{0}=x^\l$ into Eq.\z\eqref{EC} setting the right-hand side equal to zero. Therefore, it results that:
    \begin{itemize}
    	\item If we have two distinct real roots, $\l_{1,2}= \dfrac{1}{2}\left(1-a\pm\sqrt{(a-1)^2-4 b}\right)$, the homogeneous solution is $y_{h}=c_{\l_1} x^{\l_1}+c_{\l_2} x^{\l_2}$\,; 
    	\item If we have one repeated real root, $\l=-\sqrt{b}$, then the homogeneous solution is given $y_{h}=c_{1} x^{\l} + c_{2} x^{\l}\ln{x}$\,;
    	\item If we have two complex roots, $\l_{1,2}= \dfrac{1}{2}\left(1-a\pm i\sqrt{4b-(a-1)^2}\right)$, then the homogeneous solution is $y_{h}=c_1 x^{1-a}\cos{\left[(4b-(a-1)^2)\ln{x}\right]}+c_2 x^{1-a}\sin{\left[(4b-(a-1)^2)\ln{x}\right]}$.
    \end{itemize}
	Regarding the particular solution of Eq.\z\eqref{EC}, we have the following possibilities:
	\begin{itemize}
		\item If $1/\a$ is not equal to any root of the characteristic equation, then the particular solution is $y_p(x)\propto x^{\frac{1}{\a}}$\,;
		\item If $1/\a$ is equal to a root of the characteristic equation, then the particular solution is given by $y_p(x)\propto x^{\frac{1}{\a}}\left(\ln{x}\right)^\beta$, where $\beta$ is the multiplicity of the root\z\footnote{Being Eq.\z\eqref{EC} of the second order, in this case $\b$ can be only equal to 1 or 2.}.
	\end{itemize}

	Now, let us assume $B=0$. Thus, Eq.\z\eqref{ECx} takes the following form
	\begin{equation}\label{ECxx}
		-\frac{1}{6} y(x)+\frac{A}{C}\,x y'(x)=-\frac{x^{1/\a}}{3\eps\k\c\left|C\right|^{1/\a}}\,.
	\end{equation}
	In this case, we only have one (real) solution for the characteristic equation, $\l=\dfrac{C}{6A}$\,. Therefore, we only have one homogeneous solution, $y_h=x^{\l}$ , and two possible particular solutions, given by:
	\begin{itemize}
		\item If $1/\a$ is not equal to the root of the characteristic equation, then the particular solution is $y_p(x)\propto x^{1/\a}$\,;
		\item If $1/\a$ is equal to the root of the characteristic equation, then the particular solution is given $y_p(x)\propto x^{1/\a}\ln{x}$\,.
	\end{itemize}
	${}$\\
	%%%%%%%%%%%%%%%%%%%%%%%%%%%%%%%%%%%%%%%%%%%%%%%%%%%%%%%%%%%%%%%%%
	%%%%%%%%%%%%%%%%%%%%%%%%%%%%%%%%%%%%%%%%%%%%%%%%%%%%%%%%%%%%%%%%%
	
	At this point, let us focus on a particular value of $\alpha$. It is interesting to further analyse the ansatz Eq.\z\eqref{ansatz1} in the case of $\alpha=1$, which represents the easiest generalization of Refs.\z\cite{Sotiriou:2008ya,Terrucha:2019jpm}. In this case, we have
	\begin{equation}\label{ansatz1_w_1}
		\bar{\mathcal{X}}= C\, (\kappa  \rho )^{2  }\,,\quad{\rm where}\quad C=C_1 (3 w -1)^2+C_2 \left(3 w ^2+1\right)+\frac{1}{3} C_3 \left(9w^2+6w+5\right),\,
	\end{equation} 
	and, requiring that $\bar{\mathcal{X}}\neq0\,$, we must exclude the following configurations:
	\begin{gather}\label{null1} 
		C_2=-2 C_3 \quad \land  \quad  C_1=\frac{C_3}{3}\,,\\
		3 C_1+C_2+C_3\neq 0 \quad \land   \quad  w =\frac{3 C_1-C_3\pm\sqrt{-3 C_2^2-12 C_1 C_2-8 C_3 C_2-4 C_3^2-24 C_1 C_3}}{3 \left(3 C_1+C_2+C_3\right)}\,,
		\\ 3 C_1+C_2+C_3= 0  \quad \land  \quad C_2+2 C_3\neq 0  \quad  \land  \quad w =-\frac{1}{3}\,.
	\end{gather}
	In particular, if the first condition holds, the constant $C$ in Eq.\z\eqref{ansatz1_w_1} is identically zero for any value of $w$, if $3C_1+C_2+C_3=0$ we have $C=2 \left(C_2+2 C_3\right) (3 w +1)$, and for $w=-1/3$ we get $C=4(3C_1+C_2+C_3)$.

	Then, the equation for $y(x)$ results in:
	\be\label{de}
	a_0\,y(x)+a_1\, x\,y'(x)+a_2\,x^2\,y''(x)=\frac{x}{\eps \kappa \rho _c}\,,\\
	\ee
	where
	\bea
		a_0&=&\frac{1}{6}\left(C_2 \left(9 w ^2+3\right)+C_3 \left(9 w ^2+6 w +5\right)+3 C_1 (1-3 w )^2\right)\,,\\
		a_1&=&\frac{1}{3}  \left(3 C_2 \left(3 w ^2+3 w -2\right)+C_3 \left(9 w ^2+6 w -7\right)+3 C_1 (3 w -1) (3 w +5)\right)\,,\\
		a_2&=&4  (w +1) \left(3 C_1 (3 w -1)+3 \left(C_2+C_3\right) w +C_3\right)\,.
	\eea
	Again, we have to distinguish the case where the coefficient of the second order term is different from zero, $a_2\neq0$, from the $a_2=0$ case. The latter can only be realized by setting $\{3 C_1+C_2+C_3\neq 0\,\land\,w=\frac{3 C_1-C_3}{3 \left(3 C_1+C_2+C_3\right)}\}$, excluding configurations of parameters such that $\{C_2=-2 C_3\,\land\,C_1=C_3/3\}$ which implies $C=0$ in Eq.\z\eqref{redX}, and $w=-1$. The former is more interesting because it is the case of $f(R)$ and $f(G)$. Therefore, let us focus on it.
	
	From the previous discussion of Eq.\z\eqref{EC}, we already know that there are three possible particular solutions for our problem: $y_{p1}(x)\propto x$, $y_{p2}(x)\propto x\ln{x}$ and $y_{p3}(x)\propto x(\ln{x})^2$. The easiest way to obtain the proportionality constant related to each particular solution is by substituting those prototypes of solutions in Eq.\z\eqref{de}. This procedure allows us to outline the following classification:
	\begin{itemize}
		\item  For $3 C_1+C_2+C_3\neq0$ and $w\neq1/3$, we get:
		\begin{equation}
			y_{p1}(x)=\dfrac{2 x}{3 \left(3 C_1+C_2+C_3\right) \left(1+w\right)\left(3w-1\right) \epsilon \kappa \rho _c}\,.
	    \end{equation}
        The complete solution can be written in the following general form
        \begin{equation}
        	\f({\mathcal{X}})=\frac{2 \left|\mathcal{X}\right|}{3 \left(3 C_1+C_2+C_3\right) \left(1+  w\right)\left(3w-1\right)\epsilon \kappa \rho _c}+A\,\left|\mathcal{X}\right|^{(-b-\sqrt{\Delta})/(2a)}+B\,\left|\mathcal{X}\right|^{(-b+\sqrt{\Delta})/(2a)}\,,
        \end{equation}
        where $A$ and $B$ are integration constants, while $a$, $b$, and $\Delta$ are given by, respectively,
        \bea
        a&=&12 (w +1) \left(C_1 (9 w -3)+3 \left(C_2+C_3\right) w +C_3\right)\,,\\
        b&=&-3 C_1 (3 w -1) (9 w +7)-3 C_2 (9 w  (w +1)+2)-C_3 (3 w  (9 w +14)+19)\,,\\
        \Delta&=& 9 C_1^2 (3 w -1)^2 \left(9 w ^2+78 w +73\right)+6 C_3 C_1 (3 w -1) (3 w +1) \left(9 w ^2+90 w +85\right)
        \nn\\
        &&+18 C_2 C_1 (3 w -1) \left(9 w ^3+84 w ^2+81 w +2\right)+9 C_2^2 \left(9 w ^4+90 w ^3+93 w ^2+12 w +4\right)
        \nn\\
        &&+C_3^2 \left(81 w ^4+972 w ^3+1638 w ^2+972 w +241\right)+6 C_2 C_3 \left(27 w ^4+297 w ^3+399 w ^2+147 w +26\right)\,. \nn \\
        \eea
		\item For $3 C_1+C_2+C_3=0$, $C_2+2C_3\neq0$, and $w\neq\pm1/3$, we get:
		\begin{equation}
			y_{p2.1}(x)=\frac{3x\ln{x}}{\left(C_2+2 C_3\right) (9 w +11) \epsilon \kappa \rho _c}\,.
		\end{equation}
	    In this case, the complete solution turns out to be:
	    \bea
	    \f(\mathcal{X})= \frac{3 \abs{\mathcal{X}} \ln{\abs{\mathcal{X}}}}{\left( C_2+2C_3\right) (9  w  +11) \epsilon \kappa \rho _c}+A \,\abs{\mathcal{X}}^{(3  w  +1)/[12 ( w  +1)]}+B\, |\mathcal{X}| \,,
	    \eea
	    where $A$ and $B$ are integration constants. 
	    
	    \item For $w=1/3$ and $C_2+2C_3\neq0$, the above equation reduces to
	    \begin{equation}
		y_{p2.2}(x)=\frac{3x\ln{x}}{14\left(C_2+2 C_3\right) \epsilon \kappa \rho _c}\,.
		\end{equation}
	    The complete solution is then
	    \begin{equation}
	    	\f(\mathcal{X})= \frac{3 \abs{\mathcal{X}} \ln{\abs{\mathcal{X}}}}{14 \left(C_2+2 C_3\right) \epsilon \kappa \rho _c}+A\, \abs{{\mathcal{X}}}^{1/8}+B\, \abs{\mathcal{X}}\,,
	    \end{equation}
		which is analogous to the previous solution but differs to it as here we have no definition of $C_1$ (which is the coefficient of $R^2$). Therefore, $C_2$ and $C_3$ are not proportional to each other, in general. 
	\end{itemize} 
    
	Thus, our differential equation does not admit the solution $y_{p3}(x)\propto x(\ln{x})^2$, because by substituting $y_{p3}(x)\propto x(\ln{x})^2$ in Eq.\z\eqref{de} we obtain as the only possibility $x = 0$, which is not an admitted value.

	All the above solutions generalize the results obtained in Refs.\z\cite{Sotiriou:2008ya,Terrucha:2019jpm}. The $f(R)$ bouncing cosmology belongs to the first class of solutions of $\f(\mathcal{X})$, with $3 C_1+C_2+C_3\neq0$ and $w\neq1/3$, while the $f(\mathcal{G})$ bouncing cosmology belongs to the second class of solutions of $\f(\mathcal{X})$, with $3 C_1+C_2+C_3=0$, $C_2+2C_3\neq0$, and $w\neq-1/3$.
	
	In order to be complete in our analysis, it is worth saying that in the case of $a_2=0$, \textit{i.e.}, $\{3 C_1+C_2+C_3\neq 0\,\land\,w=\frac{3 C_1-C_3}{3 \left(3 C_1+C_2+C_3\right)}\}$, we only have one solution:
	\begin{equation}
		\f(\mathcal{X})= \frac{2 \left(3 C_1+C_2+C_3\right) \abs{\mathcal{X}}}{\left(C_2+2 C_3\right) \left(12 C_1+3 C_2+2 C_3\right) \epsilon \kappa \rho _c}+A\,{\abs{\mathcal{X}}}^{1/4}\,,
	\end{equation}
	where $A$ is an integration constant, and the parameters $\{C_i\}$ cannot be chosen such that the denominator of the solution is zero.\\
	
	%%%%%%%%%
	
	Let us consider the case of fixing $w=1$ in Eq.\z\eqref{de}. It is straightforward to write the three possible solutions taking into account that the coefficient of the second order term is $a_2=8(6C_1+3C_2+4C_3)$:
	\begin{itemize}
		\item  For $6C_1+3C_2+4C_3\neq0$ and $3 C_1+C_2+C_3\neq0$, we get:
		\begin{equation}
			\f({\mathcal{X}})=\frac{\left|{\mathcal{X}}\right|}{6 \left(3 C_1+C_2+C_3\right)  \epsilon \kappa \rho _c}+A\,\left|{\mathcal{X}}\right|^{(-b-\sqrt{\Delta})/(2a)}+B\,\left|{\mathcal{X}}\right|^{(-b+\sqrt{\Delta})/(2a)}\,,
		\end{equation}
		where
		\bea
		a&=&6\left(6 C_1+3 C_2+4 C_3\right),\\
		b&=&-24 C_1-15 C_2-22 C_3,\\
		\Delta&=&360 C_1^2+396 C_2 C_1+552 C_3 C_1+117 C_2^2+244 C_3^2+336 C_2 C_3\,,
		\eea 
		with $A$ and $B$ being integration constants.
		\item For $6C_1+3C_2+4C_3\neq0$ and $3 C_1+C_2+C_3=0$ we arrive at
		\bea
		\f({\mathcal{X}})= \frac{3 \,\abs{\mathcal{X}} \ln{\abs{\mathcal{X}}}}{20\left(6 C_1+C_2\right) \epsilon \kappa \rho _c}+A\, \abs{\mathcal{X}}^{1/6}+B\,\abs{\mathcal{X}},
		\eea
		where $A$ and $B$ are integration constants.
		\item For $6C_1+3C_2+4C_3=0$ and $C_2+2 C_3\neq 0$, we deduce
		\be
		\f({\mathcal{X}})=-\frac{\abs{\mathcal{X}}}{3 \left(C_2+2 C_3\right) \epsilon \kappa \rho _c}+A\,\abs{\mathcal{X}}^{1/4}\,,
		\ee
		where $A$ is an integration constant.
	\end{itemize}
	%

	%%%%%%%%%%%%%%%%%%%%%%%%%%%%%%%%%%%%%%%%%%%%%%%%%%%%%%%%%%%%%%%%%	
	\subsection{Solution II}\label{solution2}\label{SpecificII}
	%%%%%%%%%%%%%%%%%%%%%%%%%%%%%%%%%%%%%%%%%%%%%%%%%%%%%%%%%%%%%%%%%	
	
    Let us continue the discussion by proposing a further specific solution for Eq.\z\eqref{de_FF}. By analogy to Ref.\z\cite{Barros:2019pvc}, let us assume that $\f$ has a power-law form. In particular, in Ref.\z\cite{Barros:2019pvc} the authors propose a solution of the form $\f(R,\mathcal{G})=C\abs{\mathcal{G}}^\alpha\abs{R}^\b$. Here, in order to generalize this solution, we assume the following power-law form:
	\be
	\f(R,P,Q)=C_0\, \abs{R}^{\a}P^{\b}\,Q^{\g}|C_1R^2+C_2P+C_3Q|^{\d}\label{ansatz2}\,,
	\ee
	where $\{C_i\}$, with $i=1,2,3$, and $\a,\,\b,\,\g,\,\d$ are real dimensionless parameters, while $C_0$ has dimensions such that $\f$ is proportional to an energy density\z\footnote{Because $\eps$ is a real dimensionless parameter, $R$ and $\f$ must have the same dimension so that the action is well-defined.} (as well as $R$). 
	
	Our aim is to determine the expression of the constant $C_0$ such that Eq.\z\eqref{ansatz2} is a solution for Eq.\z\eqref{de_FF}. Therefore, we want to obtain the dependence of $C_0$ on the other parameters $\{C_1,C_2,C_3,\a,\b,\g,\d\}$.	Obviously, for particular choices of the parameters we already know the value of the constant $C_0$ by taking into account the discussion of the previous ansatz and the results presented in Refs.\z\cite{Sotiriou:2008ya,Terrucha:2019jpm,Barros:2019pvc}.
	
	It is important to notice that we need to exclude cases in which the above function is identically zero. Therefore, using the reduced form of variables, Eqs.\z\eqref{Rw}\z--\z\eqref{Qw}, it possible to see we need to exclude the following configurations of parameters:
	\begin{gather}\label{null2} 
		w=\frac{1}{3}\,,\\
			C_2=-2 C_3\, \quad \land\, \quad C_1=\frac{C_3}{3}\,,
			\\3 C_1+C_2+C_3\neq 0\, \quad \land\, \quad  w =\frac{3 C_1-C_3\pm\sqrt{-3 C_2^2-12 C_1 C_2-8 C_3 C_2-4 C_3^2-24 C_1 C_3}}{3 \left(3 C_1+C_2+C_3\right)}\,,
			\\3 C_1+C_2+C_3= 0\,  \quad  \land\,  \quad C_2+2 C_3\neq 0\, \quad \land\,   \quad 
			w =-\frac{1}{3}\,.
	\end{gather}
	In particular, the first condition is related to $\bar{R}=0$, while the last three
	conditions correspond to $C_1\bar{R}^2+C_2\bar{P}+C_3\bar{Q}\neq0$, \textit{i.e.},
	$3C_1 (3 w -1)^2+3C_2 \left(3 w ^2+1\right)+ C_3 \left(9w^2+6w+5\right)\neq0$.\\
  
	Substituting Eq.\z\eqref{ansatz2} in Eq.\z\eqref{de_FF}, we get the following dimensional constraint:
	\be
	\alpha =2- 2(\beta +\gamma +\d)\,.
	\label{solIIconstraint}
	\ee
	This arises from the fact that for Eq.\z\eqref{ansatz2} be a solution of Eq.\z\eqref{de_FF} we have to impose $\r^{\a+2(\b+\g+\d)}\propto \r^2$, so that this constrains $C_0$ to have dimension $[C_0]=[\r]^{-1}$. 
	Using the above condition, the ansatz\z\eqref{ansatz2} reproduces the bouncing universe of LQC by requiring that the constant $C_0$ is given by:
	\be\label{ans1const}
	C_0=\frac{2}{\epsilon \kappa \rho _c}\frac{ 3^{\gamma +\delta } ( 3  w -1 ) ^{2 (\beta +\gamma +\delta )-1} (3  w  ^2+1)^{1-\beta }  (9  w  ^2+6  w  +5)^{1-\gamma }  | U_{\mathcal{X}} | {}^{-\delta }U_{\mathcal{X}}}{U_{\mathcal{X}}\left( U_{\b} \beta  + U_\g \gamma +U_w\right)+U_\d\delta }\,,
	\ee
	where
	\bea
		U_{\mathcal{X}}&=&\left(3 C_1 (1-3  w  )^2+3C_2 \left(3  w  ^2+1\right)+C_3 \left(9  w  ^2+6  w  +5\right)\right),\\
		U_\b&=&-6(w +1) (3 w -1) (3 w +1) \left(9 w ^2+6 w +5\right),\\
		U_\g&=&-36(w +1) (3 w -1) (3 w +1) \left(3 w ^2+1\right),\\
		U_w&=&+9 (w +1) (3 w -1) \left(9 w ^2+6 w +5\right) \left(3 w ^2+1\right),\\
		U_\d&=&-18 \left(C_2+2 C_3\right) (w +1) (3 w -1) (3 w +1) \left(3 w ^2+1\right) \left(9 w ^2+6 w +5\right).
	\eea
	The choice of the parameters must be such that the constant $C_0$ be well-defined. Therefore, we exclude all the configurations of parameters such that the denominator is identically zero, namely:
	\begin{itemize}
	\item For $\left(C_2+2 C_3\right) (3 w +1)\neq 0$, the denominator of Eq.\z\eqref{ans1const} vanishes when
		\begin{equation}
			\delta =\frac{U_{\mathcal{X}}\left(3 \left(3  w  ^2+1\right) \left(-4 \gamma  (3  w  +1)+9  w  ^2+6  w  +5\right)-2 \beta  (3  w  +1) \left(9  w  ^2+6  w  +5\right)\right)}{6 \left(C_2+2 C_3\right)(3  w  +1) \left(3  w  ^2+1\right) \left(9  w  ^2+6  w  +5\right)}\,.
		\end{equation}
		\item For $C_2+2C_3=0$ and $w\neq-1/3$, the denominator of Eq.\z\eqref{ans1const} vanishes when we have
    	\begin{equation}
    	\gamma = \frac{\left(9  w  ^2+6  w  +5\right) \left(-6 \beta   w  -2 \beta +9  w  ^2+3\right)}{12 (3  w  +1) \left(3  w  ^2+1\right)}\,.
        \end{equation}%
		\end{itemize}
	Notice that for the specific case of $w=-1/3$, the denominator of Eq.\z\eqref{ans1const} vanishes when $3C_1+C_2+C_3=0$, but this condition is in fact excluded from the beginning.\\

	However, we are mainly interested in the case $w=1$, where the constant $C_0$ takes the form
	\be
	C_0= -\frac{ 3^{\gamma +\delta -1}\,5^{1-\gamma } \left(3 C_1+3 C_2+5 C_3\right) \left| 3 C_1+3 C_2+5 C_3\right| {}^{-\delta }}{2 \epsilon \kappa \rho _c \left(\left(3 C_1+3 C_2+5 C_3\right) (10 \beta +12 \gamma -15)+30 \left(C_2+2 C_3\right) \delta\right)}\,.
	\ee
	
	To conclude the discussion of this case, it is worth saying that by removing the dependence on $R$ in the ansatz,\textit{ i.e.} $\f(R,P,Q)=C_0\,{P}^{\b}{Q}^{\g}|C_2P+C_3Q|^{\d}$, and fixing $w=1/3$, it is possible to see that there is no value of $C_0$ such that the power-law form of $\f(P,Q)$ satisfies Eq.\z\eqref{de_FF}.

	%%%%%%%%%%%%%%%%%%%%%%%%%%%%%%%%%%%%%%%%%%%%%%%%%%%%%%%%%%%%%%%%%	
	\subsection{Solution III}\label{solution3}\label{SpecificIII}
	%%%%%%%%%%%%%%%%%%%%%%%%%%%%%%%%%%%%%%%%%%%%%%%%%%%%%%%%%%%%%%%%%	
	
	Finally, as last ansatz, we introduce a logarithmic dependence, as done in Ref.\z\cite{Barros:2019pvc} where $\f(R,\mathcal{G})=\mathcal{G}\ln(\abs{R}^\a\abs{\mathcal{G}}^\b)$. The reason of the latter assumption comes from the forms of particular solutions obtained in Refs.\z\cite{Sotiriou:2008ya,Terrucha:2019jpm}, which have been generalized in Subsec.\z\ref{solution1}. In particular, we propose the following solution:
	\be
	\f(R,P,Q)=C_0\, \abs{R}{}^{\a_1}\,P{}^{\b_1}\,Q{}^{\g_1}\,|C_1R^2+C_2P+C_3Q|{}^{\d_1}\ln{\left(C_{\r}\,|{R}|{}^{\a_2}\,{P}{}^{\b_2}\,{Q}{}^{\g_2}\,|C_4{R}^2+C_5{P}+C_6{Q}|{}^{\d_2}\right)}\label{ansatz3}\,,
	\ee
	where $\{C_i,\a_j,\b_j,\g_j,\d_j\}$, with $i=1,2,3$ and $j=1,2$, are real parameters, while the constant $C_{\r}$ is used to render dimensionless the argument of the logarithm\,\footnote{The constant $C_\r$ has dimension $[C_{\r}]=[\k\r]^{-\a_2-2(\b_2-\g_2-\d_2)}=[\r]^{-\a_2-2(\b_2+\g_2+\d_2)}$; one could assume $C_{\r}=(\k\c)^{-\a_2-2(\b_2+\g_2+\d_2)}$, for instance.}.
	
	As in the previous case, we get the non-vanishing conditions of the solution by considering the reduced form of the variables, Eqs.\z\eqref{Rw}\z--\z\eqref{Qw}. Therefore, we consider configurations such that
	\begin{gather}\label{null3}
		w\neq1/3\,,\\
		3(1-3 w )^2C_1+3\left(3 w ^2+1\right)C_2+(9 w ^2+6 w +5)C_3\neq 0\,,\\
		3(1-3 w )^2C_4+3\left(3 w ^2+1\right)C_5+(9 w ^2+6 w +5)C_6\neq 0\,,
	\end{gather}
    where the first one is the non-vanishing condition of $\bar{R}$, while the second and the third ones are the non-vanishing conditions of $|C_1\bar{R}^2+C_2\bar{P}+C_3\bar{Q}|$ and $|C_4\bar{R}^2+C_5\bar{P}+C_6\bar{Q}|$, respectively.

	When we replace Eq.\z\eqref{ansatz3} in Eq.\z\eqref{de_FF}, we also obtain a dimensional constraint $\alpha_1 =2- 2(\beta_1 +\gamma_1 +\d_1)$. Moreover, for Eq.\z\eqref{ansatz3} to be a solution of Eq.\z\eqref{de_FF}, we need to impose the following condition:
	\bea\label{log}
		&\left[C_2 \left(9 \omega ^2+3\right)+C_3 \left(9 \omega ^2+6 \omega +5\right)+3 C_1 (1-3 \omega )^2\right]\times&\nn\\
		&\times\left[-2 \beta _1 (3 \omega +1) \left(9 \omega ^2+6 \omega +5\right)-12 \gamma _1 (3 \omega +1) \left(3 \omega ^2+1\right)+3 \left(3 \omega ^2+1\right) \left(9 \omega ^2+6 \omega +5\right)\right]&\\
		&-6 \left(C_2+2 C_3\right) \delta _1 (3 \omega +1) \left(3 \omega ^2+1\right) \left(9 \omega ^2+6 \omega +5\right)=0\,.&\nn
	\eea
	The above condition comes from the substitution of Eq.\z\eqref{ansatz3} in Eq.\z\eqref{de_FF} and guarantees $C_0$ is a constant.
	For this specific case, from Eq.\z\eqref{log}, we can distinguish the following two cases:
	\begin{itemize}
		\item For $C_2+2C_3\neq0$ and $w\neq-1/3$, we have
		\begin{equation}
			\delta _1=\frac{  U_{\mathcal{X}1}\left(\left(9  w  ^2+6  w  +5\right) \left(-2 \beta _1 (3  w  +1)+9  w  ^2+3\right)-12 \gamma _1 (3  w  +1) \left(3  w  ^2+1\right)\right)}{6 \left(C_2+2 C_3\right)(3  w  +1) \left(3  w  ^2+1\right) \left(9  w  ^2+6  w  +5\right)}\,,
		\end{equation}
		where $U_{\mathcal{X}1}=3 (1-3  w  )^2 C_1+3 \left(3  w  ^2+1\right)C_2+ \left(9  w  ^2+6  w  +5\right)C_3$.
		\item For $C_2+2C_3=0$ and $w\neq-1/3$, we have
		\begin{equation}
			\gamma _1= \frac{\left(9  w  ^2+6  w  +5\right) \left( 3(3  w  ^2+1)-2(3w -1) \beta_1\right)}
			{12 (3  w  +1) \left(3  w  ^2+1\right)}\,.
		\end{equation}
	\end{itemize}	
	Actually, there is another case where Eq.\z\eqref{log} is satisfied, namely, for $w=-1/3$ which corresponds to $\bar{\mathcal{G}}=0$, as seen in Sec.\z\ref{sectionFLRW}. However, we will see this value corresponds to a constant $C_0=0$ and therefore it is not an acceptable value for $w$.
	
	When one of the above conditions holds, it is possible to see that the ansatz\z\eqref{ansatz3} reproduces the bouncing universe of LQC by requiring that the constant $C_0$ reads
	\begin{equation}
		C_0=-\frac{2\,3^{\gamma _1+\delta _1} (3  w  +1)(1-3  w  )^{2 (\beta _1+ \gamma _1+ \delta _1)-1} \left(3  w  ^2+1\right)^{1-\beta _1}  \left(9  w  ^2+6  w  +5\right)^{1-\gamma _1}U_{\mathcal{X}2} |U_{\mathcal{X}1}| {}^{-\delta _1} }{\epsilon \kappa \rho _c \left(U_{\mathcal{X}2} \left(U_{\a}\alpha _2 + U_{\b}\beta _2 + U_{\g} \gamma _2 \right)+U_{\d} \delta _2 \right)} \,,
	\end{equation}
	where
	\bea
		U_{\mathcal{X}1}&=&3 (1-3  w  )^2 C_1+3 \left(3  w  ^2+1\right)C_2+ \left(9  w  ^2+6  w  +5\right)C_3\,,\\
		U_{\mathcal{X}2}&=&3  (1-3  w  )^2C_4+3 \left(3  w  ^2+1\right)C_5+ \left(9  w  ^2+6  w  +5\right)C_6\,,\\
        	U_{\a}&=&\left(3  w  ^2+1\right) \left(9  w  ^2+6  w  +5\right) \left(27  w  ^2+30  w  -1\right)\,,\\
		U_{\b}&=&2(3  w  -1) \left(9  w  ^2+6  w  +5\right) \left(27  w  ^3+30  w  ^2+3  w  +4\right),\\
		U_{\g}&=&2(3  w  -1) \left(3  w  ^2+1\right) \left(81  w  ^3+117  w  ^2+51  w  +23\right),\\
		U_{\d}&=&2(3  w  -1) \left(3  w  ^2+1\right) \left(9  w  ^2+6  w  +5\right)\times
		\nn\\
		\times\left[3(3 w -1) \left(27 w ^2+30 w-1\right)\right.&C_4&\left.+3\left(27 w ^3+30 w ^2+3 w +4\right) C_5 + \left(81 w ^3+117 w ^2+51 w +23\right)C_6\right].\quad
	\eea
	For the sake of brevity, we omit writing all the configurations of the parameters which renders the denominator zero, and therefore, which must be excluded.
    
    To complete the discussion of this case, let us remove the dependence on $R$ in Eq.\z\eqref{ansatz3}, \textit{i.e.}, $\f(P,Q)=C_0\, P{}^{\b_1}\,Q{}^{\g_1}\,|C_2P+C_3Q|{}^{\d_1}\ln{\left(\tilde{P}{}^{\b_2}\,\tilde{Q}{}^{\g_2}\,|C_5\tilde{P}+C_6\tilde{Q}|{}^{\d_2}\right)}$, and fix $w=1/3$. In this way we get the following equation for $C_{0}$:
	\begin{equation}
		C_0= \frac{3\ 2^{\beta _1+\delta _1-2} \left| C_2+2 C_3\right| {}^{-\delta _1}}{7 \epsilon \kappa \rho _c \left(\beta _2+\gamma _2+\delta _2\right)}\,,
	\end{equation}
    where the dimensional constraint $\b_1+\g_1+\d_1=2$ must be satisfied.\\

    Finally, let us focus on the case $w=1$. The following condition has to be satisfied in order to guarantee that $C_0$ is constant and, therefore, in order to get a cosmological bounce:
	\be
	\left(3C_1+3C_2+5 C_3\right) \left(10 \beta _1+12 \gamma _1-15\right)+30 \left(C_2+2 C_3\right) \delta _1=0\,.
	\ee
    From the last equation, we can distinguish two cases:
			\begin{itemize}
		
		\item For $C_2+2C_3\neq 0$, we have 
		\begin{equation}
			\delta _1= -\dfrac{\left(3 C_1+3 C_2+5 C_3\right) \left(10 \beta _1+12 \gamma_1-15\right)}
			{30 \left(C_2+2 C_3\right)}\label{ans3.}\,.
		\end{equation}
		
		\item For $C_2+2C_3=0$, we have
		\begin{equation}
			\gamma _1= -\dfrac{5}{12} \left(2 \beta _1-3\right) .
		\end{equation}
		
	    \end{itemize}	
	Therefore, 	when one of the above conditions holds, it is possible to see that the ansatz in Eq.\z\eqref{ansatz3} reproduces the bouncing universe of LQC, by requiring that (for $w=1$) the constant $C_0$ is given by
	\bea
	C_0= \frac{5^{1-\gamma _1} \left(3 C_4+3 C_5+5 C_6\right) 3^{\gamma _1+\d_1} \left| 3 C_1+3 C_2+5 C_3\right| {}^{-\d_1}}{2 \epsilon \kappa \rho _c \left(\left(3 C_4+3 C_5+5 C_6\right) \left(35 \alpha _2+40 \beta _2+34 \gamma _2\right)+10 \left(21 C_4+12 C_5+17 C_6\right) \delta _2\right)}\,,
	\eea
    where it is necessary to exclude configurations of parameters which correspond to a vanishing denominator.
    
	%%%%%%%%%%%%%%%%%%%%%%%%%%%%%%%%%%%%%%%%%%%%%%%%%%%%%%%%%%%%%%%%%	
	\section{Discussion and conclusions}\label{conclusions}
	%%%%%%%%%%%%%%%%%%%%%%%%%%%%%%%%%%%%%%%%%%%%%%%%%%%%%%%%%%%%%%%%%
	
	In this paper, we found specific cosmological  models, coming from $f(R,P,Q)$ modified theory of gravity, reproducing the effective Friedmann equation of LQC. To obtain the result, we applied an order reduction technique\z\cite{Bel:1985zz,Simon:1990ic,Simon:1991bm,Sotiriou:2008ya,Terrucha:2019jpm,Barros:2019pvc}, which  consists in rewriting $f(R,P,Q)\rightarrow R+\eps\,\f(R,P,Q)$, and in expressing the geometric variables, $R$, $P$ and $Q$, in terms of energy-matter fields (at the zeroth perturbation order). This approach allows one to find additive contributions to the Ricci scalar such that the theory is perturbatively close to GR. These terms can be seen as GR corrections at the energy scales associated to the early universe, that is close to the Planck scale.

    It is worth noticing that  our approach is  semi-classical, and those terms become important as we get closer to the bounce - but not too much, i.e.,  $\rho\sim 0.1\rho_c$. In fact, in correspondence of the bounce, the small $\epsilon$ approximation is no longer valid, and the reduction of the order can no longer be performed in the way  discussed above. Namely, corrective terms are more important (but not predominate) as we get closer to the bounce where the above treatment fails. 
    By writing $f(R,P,Q)=R+\epsilon\,\varphi(R,P,Q)$ with small $\epsilon$, we consider all possible corrections to the Einstein--Hilbert action in a regime where the contribution of $R$ dominates over the small corrections. Such deviations from GR characterize the bounce when $R\ll\rho_c\sim l_P{}^{-2}$.
    In summary, the approach is based on two main ideas: 
    $(1)$ Assuming that GR is an incomplete theory, we want to extend semi-classically the Einstein's theory considering deviations from GR  due to high order curvature effects - therefore any new phenomenology is still a gravitational effect mediated by the curvature. $(2)$ Since we want to deal with metric $f(R,P,Q)$ theory of gravity, not as an exact but as an effective field theory, whose solution ought to be perturbatively close to GR, we have to perform an order reduction to the field equations such to do away with the spurious degrees of freedom. The result is an effective Lagrangian which can mimic the effective Friedmann equation of LQC far enough from the bounce. In this way, the search for bouncing solutions which are perturbatively close to GR can lead us to solutions that can be trusted reasonably close to the bounce. On the other hand, the effective Friedmann equation of LQC, from loop quantum gravity, is supposed correct at the bounce itself. We rely on the essential features of a fundamental theory, yet to be formulated,  viewed as an effective theory. In the region where the expansion is no more valid, too close to the bounce (i.e., $R\sim\epsilon\,\varphi$), one can invoke  loop quantum gravity solutions. The final fundamental theory, when formulated, would give the features of the bouncing solution described by our Lagrangian and described by loop quantum gravity at the bounce.

    The analysis carried out in this work generalizes the results obtained in previous works for $f(R)$, $f({\cal G})$ and $f(R,{\cal G})$ modified theories of gravity\z\cite{Sotiriou:2008ya,Terrucha:2019jpm,Barros:2019pvc}. It is easy to see that results we have obtained can be reduces to those  obtained in the previous works. 
    However, it is also possible to further generalize the approaches in Refs.\z\cite{Sotiriou:2008ya,Terrucha:2019jpm,Barros:2019pvc}, by considering other modified theories of gravity, in the context of \av modified-LQC Friedmann equations\cv\z\cite{Li:2019ipm}, in an analogous manner as carried out in Ref.\z\cite{Ribeiro:2021gds}, in order to further extend the analysis carried out here.
    
    At this point, it is also important to make some remark on the problem of ghosts that characterize $f(R,P,Q)$ theories. As emphasized in Ref.\z\cite{Stelle:1976gc}, for the specific case of quadratic curvature invariants in 4-dimensional action, there are eight dynamical degrees of freedom of the metric tensor. Two of these excitations are associated to the standard massless spin-2 graviton, five of them are related to a massive spin-2 particle, and the last corresponds to a massive scalar particle. Since the massive spin-2 particle has a negative definite linearized energy, then it is usually interpreted as a ghost of the theory (see Refs.\z\cite{Nunez:2004ts,Chiba:2005nz,Bogdanos:2009tn}). 
	One can assume that some still unknown mechanism (such as effects due to extra-dimensions) modify the theory so that ghosts do not appear at cosmological time scales\z\cite{Carroll:2004de}. It is also possible to consider quadratic curvature terms in the action with infinite covariant derivatives in such a way that it results ghost-free when the linearized problem is taken into account\z\cite{Buoninfante:2018xiw}.
	However, our case is quite different than the general $f(R,P,Q)$ theories, in the sense that we performed a perturbative approach to the Einstein--Hilbert action, in proximity to the Planck scale, by parameterizing $f(R,P,Q)=R +\eps \f(R,P,Q)$, with $\eps \f\ll R$. This gives us the possibility to not consider the problems related to the presence of instabilities and ghosts.
    
    Finally, it is also important to stress that, in the present work, we did not consider the most general effective action for $f(R,P,Q)$ gravity leading to bouncing cosmologies, which would be indeed technically very challenging. 
    
    In a future work, we aim to extend the present analysis to the general context of modified loop quantum cosmology.
	
	%%%%%%%%%%%%%%%%%%%%%%%%%%%%%%%%%%%%%%%%%%%%%%%%%%%%%%%%%%%%%%%%%
	\section*{Acknowledgements}
	%%%%%%%%%%%%%%%%%%%%%%%%%%%%%%%%%%%%%%%%%%%%%%%%%%%%%%%%%%%%%%%%%
	 MM, DV, and SC acknowledge support by the {\it Istituto Nazionale di Fisica Nucleare} (INFN) ({\it iniziative specifiche} MOONLIGHT-2,  TEONGRAV, and QGSKY). FSNL acknowledges support from the Fundac\~{a}o para a Ci\^{e}ncia e a Tecnologia (FCT) Scientific Employment Stimulus contract with reference CEECINST/00032/2018, and funding from the research grants No. UID/FIS/04434/2020, No. PTDC/FIS-OUT/29048/2017 and No. CERN/FIS-PAR/0037/2019. 
	%%%%%%%%%%%%%%%%%%%%%%%%%%%%%%%%%%%%%%%%%%%%%%%%%%%%%%%%%%%%%%%%%

%

\end{document}